\documentclass[11pt]{article}

\usepackage{amsmath,amsfonts,amssymb,epsfig}
\usepackage{color,epsf}
\usepackage{latexsym}
\usepackage{graphicx,stix}
\usepackage{subfig}
\usepackage{alltt}
\usepackage{orcidlink}

\newcommand{\e}{\ensuremath{\mathrm{e}}}

\newcommand{\cL}{{\mathcal L}}

\begin{document}
\title{ Approximating exponentials of commutators by optimized product formulas}

\author{F. Casas$^{1 \orcidlink{0000-0002-6445-279X}}$, A. Escorihuela-Tom\`as$^{2 \orcidlink{0000-0003-4409-3272}}$, P. A. Moreno Casares$^{3 \orcidlink{0000-0001-5500-9115}}$  \\[2ex]
$^{1}$ {\small\it Departament de Matem\`atiques and IMAC, Universitat Jaume I, 12071-Castell\'on, Spain}\\{
\small\it Email: Fernando.Casas@mat.uji.es}\\
{\small\it Corresponding author}\\[1ex]
$^{2}$ {\small\it Departament de Matem\`atiques and IMAC, Universitat Jaume I, 12071-Castell\'on, Spain}\\{
\small\it Email: alescori@uji.es}\\[1ex]
$^{3}$ {\small\it  Email: pabloantoniomorenocasares@gmail.com}\\[1ex]
}

\maketitle

\begin{abstract}

Trotter product formulas constitute a cornerstone quantum Hamiltonian simulation technique. However, the efficient implementation of Hamiltonian evolution of nested commutators remains an under explored area.
In this work, we construct optimized product formulas of orders 3 to 6 approximating the exponential of a commutator of two arbitrary operators in terms of the exponentials of the operators involved. The new schemes require a reduced number of exponentials and thus provide more efficient approximations than other previously
published alternatives. They can also be used as basic methods in recursive procedures to increase
the order of approximation.
We expect this research will improve the efficiency of quantum control protocols, as well as quantum algorithms such as the Zassenhaus-based product formula, Magnus operator-based time-dependent simulation, and product formula schemes with modified potentials.


\end{abstract}

\bigskip


\section{Introduction} \label{sec.1}

The problem of approximating the exponential of the sum of two non-commuting operators $A$ and $B$ in terms of the exponential of each operator
has a long history, with applications in many areas of physics, chemistry and applied mathematics \cite{blanes24smf}. The simplest approximation corresponds of course
to the so-called \emph{Lie--Trotter} formula $\chi^{[1]}_t = \e^{t A} \, \e^{t B}$, verifying
\begin{equation} \label{LT}
   \e^{t (A + B)} - \chi^{[1]}_t = \mathcal{O}(t^2) \qquad \mbox{ as } \; t \rightarrow 0.
\end{equation}
In the particular realm of quantum simulation, this allows one to approximate the evolution of a Hamiltonian separable into two parts, $H = H_1 + H_2$, as the product of
the evolution operator of each part, although the order of the approximation is only one. Higher orders
can be achieved by considering a symmetrized version of \eqref{LT}, also called the \emph{Strang}  \emph{scheme} (and second order Trotter product formula in quantum literature)
\begin{equation} \label{Strang}
  S^{[2]}_t = \e^{\frac{1}{2} t A} \, \e^{t B} \, \e^{\frac{1}{2} t A}.
\end{equation}
Specifically, $\e^{t (A + B)} - S^{[2]}_t = \mathcal{O}(t^3)$.  In fact, 
$S^{[2]}_t$ can be used as the starting point of the recursion ($k \ge 2$) \cite{creutz89hoh,yoshida90coh}
\begin{equation} \label{yoshida}
  S^{[2k]}_t = S^{[2k-2]}_{\gamma_k t} \, S^{[2k-2]}_{(1-2\gamma_k) t} \, S^{[2k-2]}_{\gamma_k t}, \quad \mbox{ with } \quad \gamma_k = \frac{1}{2- 2^{1/(2k-1)}},
\end{equation}
leading to approximations of arbitrarily high order $2k$. The problem, though, is that the number of required elementary exponentials increases rapidly with the order, and that
large positive and negative coefficients  appear in \eqref{yoshida}. The alternative presented in \cite{suzuki90fdo} (the so-called Suzuki--Trotter product formula)
\begin{equation} \label{suzuki}
  S^{[2k]}_t = S^{[2k-2]}_{\alpha_k t} \, S^{[2k-2]}_{\alpha_k t} \, S^{[2k-2]}_{(1-4\alpha_k) t} \, S^{[2k-2]}_{\alpha_k t} \, S^{[2k-2]}_{\alpha_k t}, \qquad 
  \alpha_k = \frac{1}{4- 4^{1/(2k-1)}},
\end{equation}
although leading to smaller error terms, involves an even larger number of exponentials. For this reason, other high-order product formulas requiring a reduced number of
exponentials have been designed along the years. They are of the form
\begin{equation} \label{splitting}
    S_t = \e^{a_1 t A} \, \e^{b_1 t B} \,  \cdots \, \e^{a_s t A} \, \e^{b_s t B} 
\end{equation}
and may be seen as a composition of the Lie--Trotter formula $\chi_t^{[1]}$ (and its adjoint) with appropriate weights. Specifically,  the coefficients $a_j$, $b_j$ in
\eqref{splitting}) are chosen so that
\[
   \e^{t (A + B)} - S_t = \mathcal{O}(t^{r+1}), 
\]
in which case $S_t$ is said to be of order $r$. This requires first obtaining and then
solving the so-called order conditions, polynomial equations in $a_j$, $b_j$ whose number and complexity grow rapidly with $r$. Thus, for $r = 1, \ldots, 10$, the number
of order conditions is, respectively, $2$, $1$, $2$, $3$, $6$, $9$, $18$, $30$, $56$ and $99$ \cite{mclachlan02sm}. 
In any case, the number of exponentials involved in \eqref{splitting} is much smaller than in compositions
\eqref{yoshida} and \eqref{suzuki} for a given order of approximation.

Formulas of this type are widely used  for the numerical integration of differential equations, 
in which case $A$ and $B$ correspond to Lie derivatives associated to different parts of the defining vector field \cite{blanes16aci}. They are called \emph{splitting methods}
in this context and possess remarkable properties related the preservation of qualitative features of the continuous system \cite{sanz-serna97gi,mclachlan02sm,hairer06gni}.
Thus, they inherit the symplectic character of the exact solution when they are applied to Hamiltonian systems in classical mechanics, and provide by construction unitary
approximations when solving the time-dependent Schr\"odinger equation. The reader is referred to the recent review \cite{blanes24smf} for a detailed
treatment and further applications, ranging from Hamiltonian Monte Carlo sampling techniques to time evolution problems in plasma physics. 
Methods for the form \eqref{splitting} are particularly appropriate for simulating the time evolution of a quantum system when the Hamiltonian is the sum over many local
interactions, so that each piece one can be efficiently evaluated by quantum circuits \cite{childs18ttf}.

Important as it is, not only the object $\e^{t(A+B)}$ can be approximated in this way. In fact, as shown in particular in \cite{jean97eao,childs13pff,chen22epf}, similar compositions can
be designed to approximate exponentials of commutators and more general Lie polynomials. The simplest compositions correspond to 
\begin{equation} \label{or.2}
  U_{2,1}(t) = \e^{t A} \, \e^{t B} \, \e^{-t A} \, \e^{-t B} \qquad \mbox{ and } \qquad   U_{2,2}(t) = \e^{-t B} \, \e^{t A} \, \e^{t B} \, \e^{-t A}, 
\end{equation}
both verifying that 
\[
   U_{2,j}(t) =  \e^{t^2 [A,B]} + \mathcal{O}(t^3), \qquad j=1,2.
\]   
Some important applications where simulating commutators are required include (i) effective Hamiltonians that contain coupling terms of this form, e.g.~\cite{childs13pff,chen22epf}, (ii)
the use of `corrector terms' to improve the accuracy of product formulas~\cite{bagherimehrab24faq} (also called modified potentials in the geometric numerical integration literature)  and (iii) Magnus operator-based approximations to the evolution of explicitly time-dependent Hamiltonian systems.

The problem of constructing product formulas of a given order $r \ge 3$, namely, determining coefficients $c_0, c_1, \ldots$ so that
\begin{equation} \label{co1}
  U_r(t) = \e^{c_0 t A} \, \e^{c_1 t B} \, \e^{c_2 t A} \, \e^{c_3 t B} \, \cdots = \e^{t^2 [A,B]} + \mathcal{O}(t^{r+1})
\end{equation}
has been analyzed in \cite{childs13pff,chen22epf}, where general recursions are presented that allow one, in principle, to achieve an arbitrary order of approximation starting
from a basic low-order product formula, in the spirit of \eqref{yoshida} and \eqref{suzuki}. The drawback of this approach, as in the previous case, is the exceedingly large
number of exponentials involved. Thus, starting with a product formula of order 3 with 6 exponentials, the most efficient product formula of order 5 involves 56 exponentials,
whereas up to 280 exponentials are required to achieve order 6. But, as it is shown in \cite{childs13pff} and especially in \cite{chen22epf}, having efficient product formulas approximating $\e^{t^2 [A,B]}$ is essential for 
some quantum simulations, in which case the exponentials $\e^{A s_1}$ and $\e^{B s_2}$ are native gates in a quantum circuit: by reducing the number of exponentials to
achieve a given order of approximation leads to a reduction of the complexity of the circuit. 

In this work we address the general problem \eqref{co1} by applying the same strategy usually pursued to construct splitting methods to approximate $\e^{t (A + B)}$: first, we
obtain the relevant order conditions required by such composition and then determine the coefficients by solving these order conditions. 
Although different procedures can be used to
get the order conditions (see for instance \cite{blanes24smf}), here we successively apply the Baker--Campbell--Hausdorff formula to the product \eqref{co1} to express
$U_r(t)$ as the exponential of only one operator (a linear combination of $A$, $B$ and their nested commutators). Comparison with $\e^{t^2 [A,B]}$ then gives the
relevant equations to be satisfied by the coefficients $c_j$.
The relative efficiency of different solutions (i.e., different sets of
coefficients) is then determined by defining a function related with the effective error, so that the most efficient scheme is taken as the one minimizing this function. 
By following this procedure, we construct
schemes of orders 4, 5 and 6 involving 10, 16 and 26 exponentials, respectively. 
Furthermore, by considering additional exponentials (and therefore additional free parameters) it is possible to get
even more efficient approximations. Of course, the methods we obtain here can be safely used as basic formulas for the recursive procedures presented in \cite{chen22epf} if
one is interested in achieving even higher orders.

Our results are summarized in Table \ref{table.1a}, where we collect the number of exponentials involved in the different product formulas presented here in comparison with
the most efficient recurrences proposed in \cite{chen22epf}. The reduction is quite remarkable and contributes a good deal to improve the efficiency. The new methods
could then be satisfactorily applied, in particular, to the quantum simulation of counter adiabatic driving and fractional quantum Hall phases on lattices \cite{chen22epf}, as
well as in control theory of classical and quantum systems \cite{childs13pff}.  In particular, in the former setting they allow to approximate the trajectory of a system 
steered along the direction indicated by the commutator up a certain order \cite{kawski02tco}.

\begin{table}[t]
  \begin{center}
    \begin{tabular}{c|c|c}
      Order $r$ & \quad Ref. \cite{chen22epf} & \quad This work \\ 
      \hline
      3   & \quad  6 & \quad 6  \\ 
     4   & \quad \textit{30} & \quad 10 \\
      5   & \quad 56 & \quad 16 \\
      6   & \quad \textit{280} & \quad 26 \\
      \hline
    \end{tabular}
    \caption{\small{Number of elementary exponentials of $A$ and $B$ involved in product formulas for approximating $\e^{t^2 [A,B]}$ up to order $r$ by applying the recurrences
    of ref. \cite{chen22epf} (second column) and those built in this paper (third column). Approximations of order 4 and 6 are not actually constructed in \cite{chen22epf}.}}
    \label{table.1a}
  \end{center}
\end{table}

The rest of this work is organized as follows. In section \ref{sec.2} we consider in detail the case of order 3, thus illustrating the technique we use and the different types of
composition that arise, whereas in section \ref{sec.3} we analyze the existing invariances and present new schemes of orders 4, 5 and 6. These methods are illustrated
in section \ref{sec.4} on several examples. Product formulas designed to approximate the exponential of other elements in the Lie algebra generated by $A$ and $B$
are presented in section \ref{sec.5}. Finally, section \ref{sec.6} contains some concluding remarks.

\section{Third-order schemes} 
\label{sec.2}

Just as it is the case for order 2, equation \eqref{or.2}, two types of compositions may in principle lead to higher order approximations, namely
\begin{eqnarray} \label{BA}
\mathrm{AB}: & \qquad & \e^{c_0 t A} \, \e^{c_1 t B} \, \cdots \, \e^{c_{2n} t A} \, \e^{c_{2n+1} t B}, \nonumber \\
\mathrm{BA}: & \qquad  & U(t) = \prod_{j=0}^n \e^{c_{2j} t B} \, \e^{c_{2j+1} t A} =  \e^{c_0 t B} \, \e^{c_1 t A} \, \cdots \, \e^{c_{2n} t B} \, \e^{c_{2n+1} t A}.
\end{eqnarray}
For our analysis next we consider compositions \eqref{BA}, although, as we will see, the results are also valid for type AB.

If we apply sequentially the Baker--Campbell--Hausdorff formula to \eqref{BA}, we end up with $U(t)$ expressed as one exponential of a series of
operators in powers of $t$, namely
\begin{equation} \label{modH}
  U(t) = \exp(Y(t)), \qquad \mbox{ with } \qquad Y(t) = \sum_{j \ge 1} t^j Y_j,
\end{equation}
and $Y_j$ is an element of the homogeneous subspace $\mathcal{L}_j(A,B)$ of degree $j$ of the graded free Lie algebra 
$\mathcal{L}(A,B) = \bigoplus_{j\geq 1} \mathcal{L}_j(A,B)$ \cite{munthe-kaas99cia}. One can think of $\mathcal{L}_n$ as the vector subspace formed by linear combinations
of nested commutators involving $n$ operators $A$ and $B$, verifying $[\mathcal{L}_n, \mathcal{L}_m] \subset \mathcal{L}_{n+m}$. For the
particular basis of $\mathcal{L}_j$ ($1 \le j \le 6$) collected in Table \ref{table.1}, the operator $Y(t)$ reads
 \begin{equation} \label{bch1}
 \begin{aligned}
   &  Y(t) = t (w_{1,1} A + w_{1,2} B) + t^2 w_{2,1} [A,B] + t^3 ( w_{3,1} E_{3,1} + w_{3,2} E_{3,2}) \\
   & \qquad + t^4 \sum_{\ell = 1}^3 w_{4,\ell} E_{4, \ell} +  t^5 \sum_{\ell = 1}^6 w_{5,\ell} E_{5, \ell}  +
    t^6 \sum_{\ell = 1}^9 w_{6,\ell} E_{6, \ell} + \mathcal{O}(t^7),
 \end{aligned}  
\end{equation}
where $w_{n,k}$ are polynomials of homogeneous degree $n$ in the parameters $c_j$. In particular,
\[
 \begin{aligned}
   & w_{1,1} = \sum_{i=0}^{2n-1} c_{2i+1}, \qquad w_{1,2} = \sum_{i=0}^{2n-1} c_{2i}, \qquad w_{12} = \frac{1}{2} w_{1,1} w_{1,2} - \sum_{i=0}^{n-1} c_{2i} \sum_{j=i}^{n-1}  c_{2j+1}. 
\end{aligned}
\]
From \eqref{bch1}, it is clear that an approximation to $\e^{t^2 [A,B]}$ of order 3 (i.e., with an error $\mathcal{O}(t^4)$) is obtained by requiring that
\begin{equation} \label{eq:wordre3}
  w_{1,1} = w_{1,2} = 0,  \quad w_{2,1} = 1, \quad w_{3,1} = w_{3,2} = 0.
\end{equation}  

\begin{table}[t]
\small
\begin{center}
  \begin{tabular}{c|lll}
    $\cL_j$ & \multicolumn{3}{c}{{\rm Basis of} $\cL_j$}   \\    \hline
$\cL_1$ & $E_{1,1}=A$ & $E_{1,2}=B$  &      \\   \hline
$\cL_2$ & $E_{2,1}=[A,B]$ &   &      \\   \hline
$\cL_3$ & $E_{3,1}=[A,E_{2,1}]$   & $E_{3,2}=[B, E_{2,1}]$  &      \\   \hline
$\cL_4$ & $E_{4,1} = [A,E_{3,1}]$  & $E_{4,2}=[B, E_{3,1}]$  &  $E_{4,3}=-[B, E_{3,2}]$      \\   \hline
$\cL_5$ & $E_{5,1}= [A, E_{4,1}]$  & $E_{5,2}=[B,E_{4,1}]$ & $E_{5,3}=[A,E_{4,2}]$   \\ 
              & $E_{5,4}=[B, E_{4,2}]$ & $E_{5,5}=[A, E_{4,3}]$ & $E_{5,6}=[B, E_{4,3}]$ \\ \hline
    $\cL_6$ & $E_{6,1}=[A, E_{5,1}]$  & $E_{6,2}=[B,E_{5,1}]$  & $E_{6,3}=[A,E_{5,2}]$  \\  
                 &  $E_{6,4}=[A, E_{5,4}]$  & $E_{6,5}=[B,E_{5,2}]$  & $E_{6,6}=[A,E_{5,5}]$  \\
                 &  $E_{6,7}=[B, E_{5,5}]$  & $E_{6,8}=[A,E_{5,6}]$  & $E_{6,9}=[B,E_{5,6}]$  \\ \hline
\end{tabular}
\caption{\small{Particular basis of $\cL_j$, $1 \le j \le 6$, taken in this work. The subspace $\cL_j$ is formed by all linear combinations
of nested commutators involving $j$ operators $A$ and $B$.} }\label{table.1}
\end{center}
\end{table}

\noindent
In consequence, at least 5 exponentials in the composition \eqref{BA} are required. It turns out, however, that equations \eqref{eq:wordre3} do not have solutions in that case. It is
therefore necessary to include an additional exponential, i.e., to consider the scheme
\begin{equation} \label{or3}
  U_3(t) =  \e^{c_0 t B} \, \e^{c_1 t A} \,  \e^{c_2 t B} \, \e^{c_3 t A} \,  \e^{c_4 t B} \, \e^{c_5 t A},
\end{equation}  
in which case one has a free parameter, say $c_5$. {The corresponding nonlinear system of equations \eqref{eq:wordre3} can be solved analytically with a computer
algebra system and the general solution reads ($c_5 \ne 0 $)}
\begin{equation}\label{eq:sol3}
  c_0 =\frac{1\mp\sqrt{5}}{2c_5}, \;\; c_1=\frac{c_5 (-1\pm\sqrt{5})}{2}, \;\; c_2=\frac{1}{c_5}, \;\;
  c_3 = \frac{c_5(-1\mp\sqrt{5})}{2}, \;\; c_4=\frac{-3\pm\sqrt{5}}{2c_5}.
\end{equation}
{We should notice that the 3rd-order scheme designed in \cite{chen22epf} is recovered by taking $c_5 = 1$.}
It is a common practice when designing splitting methods to choose the free parameter so as to minimize some objective function typically related with the 
leading error term in the asymptotic expansion of $Y(t)$ in \eqref{bch1}, in this case $\sum_{\ell=1}^3 w_{4,\ell} E_{4,\ell}$. {In absence of further information about the
particular operators $A$ and $B$ involved, the standard strategy consists in assigning the same weight to each term $E_{4,\ell}$. Thus,}
in general, given a composition \eqref{BA}
involving $s$ exponentials that approximates $\e^{t^2 [A,B]}$ up to order $r$, we define its effective error by
\begin{equation} \label{eq:ef}
   \mathcal{E}^{(r+1)} = s \ \left( \sqrt{\sum_{\ell=1}^{\dim \cL_{r+1}} |w_{r+1,\ell}|^2 } \right)^{1/r},
\end{equation}
to take also into account the computational effort required by the scheme \cite{blanes24smf}. A method of order $r$ is typically more efficient in practice 
than another one when it leads to a smaller value of $\mathcal{E}^{(r+1)}$,
so it makes sense to take the effective error as the objective function to minimize. One should be aware, however, that the value of $\mathcal{E}^{(r+1)}$
may differ depending on the  basis of $\cL_j$ one is using.

In the particular case of the 3rd-order scheme \eqref{or3}-\eqref{eq:sol3},
the minimum value of $\mathcal{E}^{(4)}$ is obtained with the solution
\begin{equation}\label{eq:sol3ncp1} 
 \begin{aligned}
   & c_0 =\pm\sqrt{\frac{2}{\sqrt{5}+1}},\qquad c_1=\mp\sqrt{\sqrt{5}-2},\qquad c_2=\mp\sqrt{\frac{2}{\sqrt{5}-1}}, \\
   & c_3 = -c_2, \qquad  c_4=-c_1,\qquad c_5=-c_0.
 \end{aligned}  
\end{equation}
Notice the peculiar arrangement in the distribution of the coefficients in this case: we can represent the resulting product formula \eqref{or3}-\eqref{eq:sol3} as
\begin{equation} \label{eq:sol3ncp} 
    \mbox{(Order 3):}  \qquad\qquad (c_0 B, \, c_1 A, \, c_2 B, \, -c_2 A, \, -c_1B, \, -c_0 A).
\end{equation}    
In fact, the same minimum is also obtained by the composition ($i = \sqrt{-1}$)
\begin{equation}\label{eq:sol3pcp} 
   (\tilde{c}_0 B, \, \tilde{c}_1 A, \, \tilde{c}_2 B, \, \tilde{c}_2 A, \, \tilde{c}_1 B, \, \tilde{c}_0 A), \qquad\qquad \tilde{c}_0 = i c_0, \qquad \tilde{c}_1 = -i c_1, \qquad \tilde{c}_2 = i c_2,
\end{equation}
with $c_j$ given in \eqref{eq:sol3ncp1}, although in this case all the coefficients are pure imaginary. Sequences such as (\ref{eq:sol3ncp}) and  \eqref{eq:sol3pcp} 
turn out to be very convenient when
designing higher order methods.

\section{Higher-order compositions}
\label{sec.3}

Before proceeding to the actual construction of higher-order schemes, it is worth to briefly review some of the invariances that a composition of type \eqref{BA}
has when approximating $\e^{t^2 [A,B]}$. This, on the one hand, provides more insight into the 
solutions obtained at order 3 and, on the other hand, are useful in our exploration.

\subsection{Invariances}
\label{sec.3.1}

The first and most obvious invariance takes place when one reverses the sign of $t$ in \eqref{BA}. If $U_r(t)$ is a method of order $r$, 
\[
   U_r(t) = \prod_{j=0}^n \e^{c_{2j} t B} \, \e^{c_{2j+1} t A}  = \exp \big(t^2 [A,B] + D_{r+1} t^{r+1} + \mathcal{O}(t^{r+2}) \big),
\]
then
\[
  U_r(-t) =  \prod_{j=0}^n \e^{-c_{2j} t B} \, \e^{-c_{2j+1} t A}  =  \exp \big(t^2 [A,B] + (-1)^{r+1} D_{r+1} t^{r+1} + \mathcal{O}(t^{r+2}) \big).
\]
In consequence, and according with the definition of $\mathcal{E}^{(r+1)}$, both $U_r(t)$ and $U_r(-t)$ lead to the same effective error. This is in agreement with
the two families of solutions \eqref{eq:sol3ncp1} obtained before and differing by a sign. In other words, \emph{when we propose a method with some specific coefficients $c_j$, 
the scheme obtained by considering the same composition with coefficients $-c_j$ has the same effective error}.

Consider now the effect of the transformation
\[
   t \longmapsto i \, \tilde{t}, \qquad A \longmapsto - \tilde{A}
\]
on composition \eqref{BA}. Then
\[
\begin{aligned}
    U_r(t) \longmapsto    \widetilde{U}_r(\tilde{t}) & =  \e^{i c_0 \tilde{t} B} \, \e^{-i c_1 \tilde{t} \tilde{A}} \, \cdots \, \e^{i c_{2n} \tilde{t} B} \, \e^{-i c_{2n+1} \tilde{t} \tilde{A}} \\
    & =  \exp \big(\tilde{t}^2 [\tilde{A},B] + (i)^{r+1} \widetilde{D}_{r+1} \, \tilde{t}^{r+1} + \mathcal{O}(\tilde{t}^{r+2}) \big),
\end{aligned}
\]   
where $\widetilde{D}_{r+1}$ is the same linear combination of commutators as $D_{r+1}$, possibly with some coefficients with the reverse sign, thus leading again to a scheme
with the same effective error. This fact accounts for the second family of solutions \eqref{eq:sol3pcp} found when $r=3$.

Finally, consider the transformation
\begin{equation} \label{tab}
   A \longmapsto \hat{B}, \qquad B \longmapsto -\hat{A}.
\end{equation}
Then, clearly
\[
\begin{aligned}
    U_r(t) \longmapsto    \hat{U}_r(t) & =  \e^{-c_0 t \hat{A}} \, \e^{c_1 t \hat{B}} \, \cdots \, \e^{-c_{2n} t \hat{A}} \, \e^{c_{2n+1} t \hat{B}} \\
    & =  \exp \big(t^2 [\hat{A}, \hat{B}] +  \widehat{D}_{r+1} \, t^{r+1} + \mathcal{O}(t^{r+2}) \big).
\end{aligned}
\]   
In other words, \emph{a BA method is transformed into an AB scheme}, although possibly with a different value of $\mathcal{E}^{(r+1)}$, since the map \eqref{tab} may modify the
basis. In the particular case of the basis collected in Table \ref{table.1}, one has $\hat{E}_{6,4} = E_{6,4} + \frac{1}{3} (E_{6,5} + E_{6,6})$. All other terms $\hat{E}_{j,\ell}$
up to $j=6$ only differ from $E_{j,\ell}$ in a sign. We can therefore safely consider schemes of type BA, at least up to this order of approximation. Notice how
$U_{2,1}(t)$ can be obtained from $U_{2,2}(t)$ in \eqref{or.2} in this way.

\subsection{`Counter-palindromic' patterns}

We have seen how compositions \eqref{eq:sol3ncp} and \eqref{eq:sol3pcp} lead, when $r=3$, to methods with the minimum value of the effective error, such as
is defined in \eqref{eq:ef}. Notice that scheme $U_{2,2}(t)$ in \eqref{or.2} actually follows the same pattern. It seems then reasonable to consider similar
sequences of coefficients when the order of approximation $r > 3$, i.e., either
\begin{equation} \label{pcp}
   (c_0, c_1, c_2, \ldots, c_{m-1}, c_m, c_m, c_{m-1}, \ldots, c_2, c_1, c_0)
\end{equation}   
or
\begin{equation} \label{ncp}
   (c_0, c_1, c_2, \ldots, c_{m-1}, c_m, -c_m, -c_{m-1}, \ldots, -c_2, -c_1, -c_0).
\end{equation}   
Schemes of this class will be called (positive or negative) \emph{counter-palindromic} compositions,
to emphasize the fact that the coefficients multiply a different operator ($A$ or $B$)
the second time they appear in the sequence (with the same or opposite sign, respectively). Sequences of this type have been previously considered
in \cite{gray96sit} when designing symplectic integrators for the (space discretized) Schr\"odinger equation.

Moreover, it turns out that sequences \eqref{pcp} and \eqref{ncp} actually
lead to a significant reduction in the number of order conditions, since they already satisfy by construction many of them. 
Specifically, the coefficients in the expansion \eqref{bch1} of the operator $Y(t)$ associated to the scheme verify the identities (up to order 6):
\[
 \begin{array}{lllc}
    w_{1,1} = \pm w_{1,2} \qquad & w_{3,1} = \mp w_{3,2} \qquad & w_{4,1} = - w_{4,3}  &  \\
    w_{5,1} = \pm w_{5,6} \qquad & w_{5,2} = \pm w_{5,5}  \qquad& w_{5,3} = \pm w_{5,4} & \\
    w_{6,1} = -w_{6,9} \qquad & w_{6,2} = -w_{6,8}  \qquad & w_{6,3} = -w_{6,7} \qquad & w_{6,4} = 3 (w_{6,5} + w_{6,6}).
 \end{array}
\]   
Here the top sign corresponds to the pattern \eqref{pcp} and the bottom sign to \eqref{ncp}.

This can be seen as follows. Suppose we have a composition of type \eqref{pcp},
\begin{equation} \label{pcp.1d}
 U(t)  =  \e^{c_0 t B} \, \e^{c_1 t A}  \, \cdots \, \e^{c_{m-1} t B} \, \e^{c_m t A} \, \e^{c_m t B} \, \e^{c_{m-1} t A} \, \cdots \, \e^{c_1 t B} \, \e^{c_0 t A},
\end{equation}
which can be formally expressed as $ U(t) = \exp (Y(t))$, with $Y(t)$ given by \eqref{bch1},
and interchange $A$ and $B$ in \eqref{pcp.1d}. This results in the new method
\[
  \widetilde{U}(t) = \e^{c_0 t A} \, \e^{c_1 t B}  \, \cdots \, \e^{c_{m-1} t A} \, \e^{c_m t B} \, \e^{c_m t A} \, \e^{c_{m-1} t B} \, \cdots \, \e^{c_1 t A} \, \e^{c_0 t B},
\]
whose associated operator $\widetilde{Y}(t)$ in $\widetilde{U}(t) = \exp (\widetilde{Y}(t))$ reads
\[
 \begin{aligned}
   &  \widetilde{Y}(t) = t (w_{1,1} B + w_{1,2} A) - t^2 w_{2,1} [A,B] + t^3 ( -w_{3,1} E_{3,2} + w_{3,2} E_{3,1}) \\
   &  \qquad + t^4 ( w_{4,1} E_{4,3} - w_{4,2} E_{4,2} + w_{4,3} E_{4,1} ) + \mathcal{O}(t^5).
 \end{aligned}  
\] 
Now, it turns out that $\big(\widetilde{U}(-t) \big)^{-1} = U(t)$, so that
\[
  - \widetilde{Y}(-t) = Y(t), 
\]
whence, by comparing the corresponding expansions, one arrives at $w_{1,1} = w_{1,2}$, $w_{3,1} = - w_{3,2}$, $w_{4,1} = - w_{4,3}$, etc. A similar argument is also valid for
compositions \eqref{ncp} by considering the interchange $A \longleftrightarrow -B$.

As a result, designing counter-palindromic methods of order $r \ge 4$ requires solving a smaller number of polynomial equations than with more general compositions. 
This is clearly visible in Table \ref{table.2}, where we collect the number of order conditions required to achieve order $3 \le r \le 6$ for a general composition \eqref{BA}
(first row) and the corresponding number for a counter-palindromic composition (second row). For clarity, we have also included the minimum number $s$ of exponentials
required in each case (in \textbf{bold}). {We remark that the number of exponentials has to be equal or greater than the number of order conditions, so that one has enough parameters to satisfy all the required equations. In other words, for an approximation of order $r$ one has
\[
     s \ge \sum_{j=1}^r \dim \cL_j.
\]     
}

\begin{table}[t]
  \begin{center}
    \begin{tabular}{c|cccc}
      Order $r$ & \quad 3 & \quad 4 & \quad 5 & \quad 6\\ 
      \hline
      General ($s$)   & \quad  5 (\textbf{6}) & \quad 8 (\textbf{$\ge$ 8}) & \quad 14 (\textbf{$\ge$ 14}) & \quad 23 (\textbf{$\ge$ 23}) \\ 
      CP ($s$)  & \quad  3 (\textbf{6}) & \quad  5 (\textbf{10}) & \quad 8 (\textbf{16}) &  \quad 13 (\textbf{26}) \\
    \end{tabular}
    \caption{\small{Number of independent order conditions to achieve order $r$ for a general composition \eqref{BA} of type BA (first row) and for counter-palindromic compositions 
    of the form \eqref{pcp}-\eqref{ncp} (second
    row). The minimum number $s$ of elementary exponentials required in each case is also included (in \textbf{bold}).}}
    \label{table.2}
  \end{center}
\end{table}

Although the number of exponentials increases with respect to the general case (e.g., 16 vs. 14 for $r=5$), the number of equations to solve (and therefore the
complexity of the problem) reduces considerably (14 vs. 8 equations for $r=5$). Moreover, although the number of order conditions sets up the minimum number of exponentials
involved in a general composition, this by itself does not guarantee the existence of real solutions. This fact has been illustrated for $r=3$ in section \ref{sec.2}. It is illustrative
to compare these results with Table \ref{table.1a}.

\subsection{New schemes}

We have explored both types of sequences \eqref{pcp} and \eqref{ncp} and solved the required order conditions to achieve up to order 6 with the minimum number 
of exponentials in each case. This task is simplified by considering schemes with patterns \eqref{pcp} and \eqref{ncp}.
The coefficients of the most efficient methods (i.e., with the smallest value of $\mathcal{E}^{(r+1)}$) are collected in Table
\ref{table.3}. The resulting method is denoted as $\mathcal{PCP}_{s}^{[r]}$ (respectively, $\mathcal{NCP}_{s}^{[r]}$) if it corresponds to a composition of class \eqref{pcp}
(respect., \eqref{ncp}) of order $r$ involving $s$ exponentials. For clarity, the methods read, respectively 

\begin{equation}   \label{eq.pcp_expl} 
    \renewcommand\arraystretch{1.1}
  \mathcal{PCP}:  \;\;  \begin{matrix} 
    U(t) = (c_0 B, \, c_1 A,  \, \cdots \, c_{m-1} B, \, c_m  A, \, c_m  B, \, c_{m-1} A, \, \cdots \, c_1 B, \, c_0 A) \quad \mbox{ ($m$ odd) } \\
     U(t) = (c_0 B, \, c_1 A,  \, \cdots \, c_{m-1} A, \, c_m  B, \, c_m  A, \, c_{m-1} B, \, \cdots \, c_1 B, \, c_0 A) \quad \mbox{ ($m$ even) } 
  \end{matrix}
\end{equation}   
and
\begin{equation}   \label{eq.ncp_expl} 
    \renewcommand\arraystretch{1.1}
  \mathcal{NCP}:  \;\;  \begin{matrix} 
    U(t) = (c_0 B, \, c_1 A,  \, \cdots \, c_{m-1} B, \, c_m  A, \, -c_m  B, \, -c_{m-1} A, \, \cdots \, -c_1 B, \, -c_0 A) \quad \mbox{ ($m$ odd) } \\
    U(t) = (c_0 B, \, c_1 A,  \, \cdots \, c_{m-1} A, \, c_m  B, \, -c_m  A, \, -c_{m-1} B, \, \cdots \, -c_1 B, \, -c_0 A)  \quad \mbox{ ($m$ even). } 
  \end{matrix}
\end{equation}   
Thus, scheme $\mathcal{PCP}_{16}^{[5]}$ in Table \ref{table.3} corresponds to the composition
\[
  (c_0 B, \, c_1 A, \,  \ldots, \, c_6 B, \, c_7 A, \, c_7 B, \, c_6 A, \, \ldots, \, c_1 B, \, c_0 A),
\] 
etc. We should remark that, due to the symmetries and invariances explored in subsection \ref{sec.3.1}, there are more compositions than those collected in Table
\ref{table.3} leading to the same efficiency.

\begin{table}[t!]
  \centering
    \renewcommand\arraystretch{1.1}
    \begin{tabular}{lll}
      \multicolumn{3}{c}{$\mathcal{NCP}_{6}^{[3]} \qquad\qquad \mathcal{E}^{(4)}/s \approx 0.473$}\\
      \hline 
      $c_0 = c_1 - c_2$ &\qquad $c_1=-\sqrt{\sqrt{5}-2}$ \\
      $c_2=-\sqrt{\frac{2}{\sqrt{5}-1}}$ & \\
      \hline 
                                    & & \\
      \multicolumn{3}{c}{$\mathcal{NCP}_{10}^{[4]} \qquad\qquad \mathcal{E}^{(5)}/s \approx 0.606$}\\
      \hline 
      $c_0 = \sum_{j=1}^4 (-1)^{j+1} c_j$ &\qquad $c_1=0.4920434066428167763156$ \\
      $c_2=-1.569846260451462851779$ & \qquad  $c_3=-0.0340560371300231615989$ \\
      $c_4=3.007307207357765662262$ & \\
      \hline 
                                    & & \\                                    
      \multicolumn{3}{c}{$\mathcal{PCP}_{16}^{[5]} \qquad\qquad \mathcal{E}^{(6)}/s \approx 0.505$}\\
      \hline 
      $c_0 = - \sum_{j=1}^7  c_j$ &\qquad $c_1=0.2969175443796203417835$ \\
      $c_2=1.418243492034305431995$ & \qquad  $c_3=0.4347212029859471608694$ \\
      $c_4=-0.127142127469064995044$ & \qquad $c_5=-2.014276365712093993010$ \\
      $c_6=0.8493401946712687892513$ & \qquad $c_7=-0.305642216160471071886$ \\
       \hline 
                                    & & \\                                    
      \multicolumn{3}{c}{$\mathcal{PCP}_{26}^{[6]} \qquad\qquad \mathcal{E}^{(7)}/s \approx 0.447$}\\
      \hline 
      $c_0 = - \sum_{j=1}^{12}  c_j$ &\qquad $c_1=0.2464427486685065253599$ \\
      $c_2=0.437855533639627516106$ & \qquad  $c_3=-0.6290554972825559401392$ \\
      $c_4=-1.160402744300525331934$ & \qquad $c_5=-0.5248160600039844378749$ \\
      $c_6=-0.2264322765760404736976$ & \qquad $c_7= 0.1165418804073705040233$ \\
      $c_8=0.4687839445292851414849$ & \qquad $c_9= 1.983312306755703005101$ \\
      $c_{10}=-0.9894918460835968618662$ & \qquad $c_{11}=0.6722571007458945095097$ \\
      $c_{12}=-0.2387711966553848135336$ & \\
      \hline 
    \end{tabular}
  \caption{\small{Coefficients of the most efficient counter-palindromic schemes of order $3 \le r \le 6$ with the minimum number of exponentials. 
  $\mathcal{PCP}_{s}^{[r]}$ refers to a composition of the form \eqref{eq.pcp_expl} of order $r$ with $s$ exponentials, whereas $\mathcal{NCP}_{s}^{[r]}$ denotes
  a method of the form \eqref{eq.ncp_expl}.
  \label{table.3}}}
\end{table}

A useful strategy to improve the efficiency of splitting methods when approximating $\e^{t(A+B)}$ consists in including additional exponentials
in the composition, and therefore additional parameters, which are then fixed in the optimization process of the objective function \cite{blanes24smf,mclachlan02sm}. In fact,
very often the most efficient method does not necessarily correspond to the scheme requiring the minimum number of exponentials: the extra cost can be
compensated by the additional accuracy obtained, although solving the polynomial equations with additional exponentials and free parameters is not a trivial
task. By pursuing the same strategy in this context we have obtained the schemes of order 4 and 5 collected in Table \ref{table.4}. Notice how the effective error is
reduced: the extra cost resulting from an increased number of exponentials is compensated by the higher accuracy achieved.

\begin{table}
  \centering
    \renewcommand\arraystretch{1.1}
    \begin{tabular}{lll}
      \multicolumn{3}{c}{$\mathcal{PCP}_{12}^{[4]} \qquad\qquad \mathcal{E}^{(5)}/s \approx 0.455$}\\
      \hline 
      $c_0 = - \sum_{j=1}^5  c_j$ &\qquad $c_1=0.3263285743794757829237$  \\
      $c_2=-1.564170317916158642032$ & \qquad  $c_3=-0.0234725141740210902965$ \\
      $c_4=2.920816850699232751348$ & \qquad $c_5=-0.8045459762846959202889$ \\
       \hline 
                                    & & \\                                    
      \multicolumn{3}{c}{$\mathcal{NCP}_{18}^{[5]} \qquad\qquad \mathcal{E}^{(6)}/s \approx 0.395$}\\
      \hline 
      $c_0 = \sum_{j=1}^8 (-1)^{j+1} c_j$ &\qquad $c_1=-0.6410115692148225407946$ \\
      $c_2=0.3165189600901244909982$ & \qquad  $c_3=0.2075766074841999769730$ \\
      $c_4=-1.042459743800714071012$ & \qquad $c_5=1.027769699504593533740$ \\
      $c_6=1.290831433928573680468$ & \qquad $c_7=0.7061407649397449413288$ \\
      $c_8=0.253358191085494126186$ &  \\
      \hline 
    \end{tabular}
  \caption{\small{Coefficients of optimized counter-palindromic schemes $\mathcal{PCP}_{12}^{[4]}$ of order 4 with 12 exponentials and $\mathcal{NCP}_{18}^{[5]}$ of order 5 with 
  18 exponentials. Same notation as in Table \ref{table.3}. \label{table.4}}}
\end{table}

\section{Numerical examples}
\label{sec.4}

\begin{figure}[!ht]
  \begin{center}
      \includegraphics[scale=0.48]{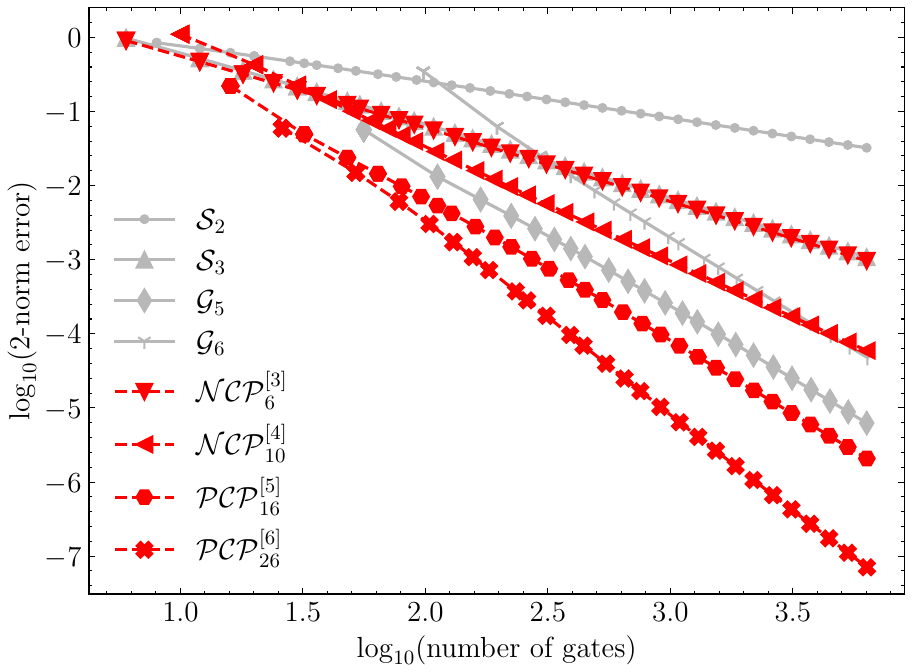}
      \includegraphics[scale=0.48]{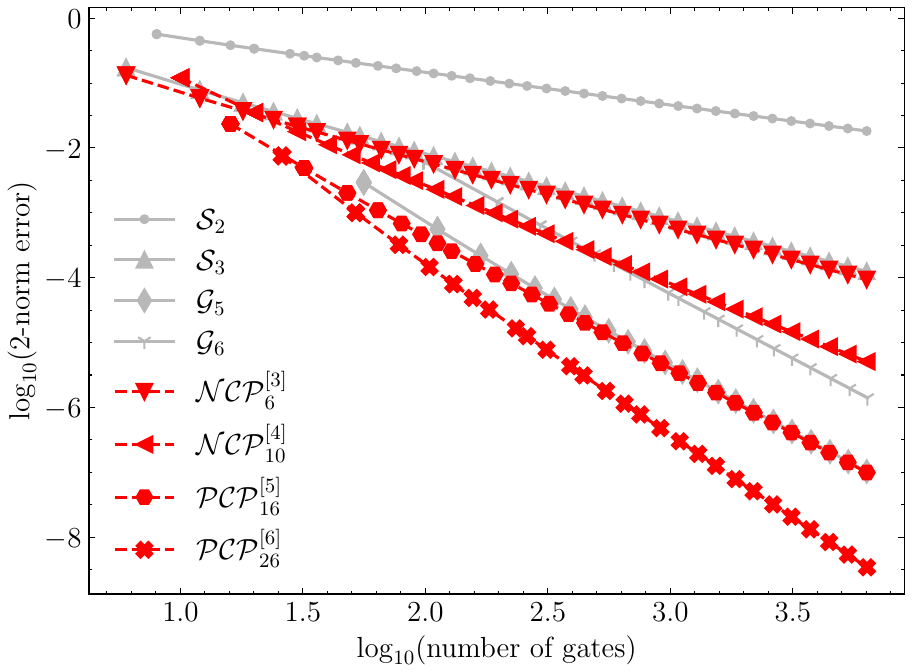}
    \caption{\small{Error committed in the approximation of $\e^{[A,B]}$ by different product formulas vs. the number of elementary exponentials when 
    $A = -i \sigma_x$, $B = -i \sigma_z$ (left) and when $A$ and $B$ are $16 \times 16$ real matrices with random elements (right). The new methods of Table \ref{table.3} (in red)
    are compared with schemes designed in \cite{chen22epf} (in gray). The improvement in efficiency of the new methods {of order $4$, $5$ and $6$} is clearly visible, 
    {whereas
    the curves corresponding to the 3rd-order schemes $\mathcal{S}_3$ and $\mathcal{NCP}_6^{[3]}$ almost coincide.}}}
    \label{fig:effic}
  \end{center}
\end{figure}

An important problem in quantum control 
consists in approximating the exponential of the commutator of Pauli matrices, namely $\e^{ [-i \sigma_x, -i \sigma_z] }$,
 by products of elementary exponentials of  $A = -i \sigma_x$ and $B = -i \sigma_z$, and 
has also been used as a test bench in \cite{chen22epf}.  
Specifically, we compute
\begin{equation} \label{err_ne}
  \big\| U(1/\sqrt{n})^n - \e^{[A,B]} \big\|_2
\end{equation}
for several values of $n$  and plot its error as a function of the total number of elementary exponentials $n \cdot s$ required by each method $U(t)$
to take into account its overall computational cost. This number is denoted as the number of gates in the graphs. The diagram of Figure \ref{fig:effic} (left) shows the results achieved
by the new schemes (in red color) in comparison with the most efficient recursions available in the literature (in gray). Different markers identify different methods.

Specifically, $\mathcal{S}_2$, $\mathcal{S}_3$ and $\mathcal{G}_5$ correspond to schemes of order 2, 3 and 5 proposed in \cite{chen22epf}, involving 4, 6 and 56 exponentials,
respectively. Actually, $\mathcal{S}_2$ is the approximation $U_{2,1}$ of \eqref{or.2}, whereas $\mathcal{G}_5$ is obtained by applying the  
($\sqrt{10}$-copy) recursive formula
presented in \cite{chen22epf} to the basic scheme $\mathcal{S}_3$. We also show in the diagram the results achieved by the methods of Table \ref{table.3}. Notice that {whereas the new 3-order approximation $\mathcal{NCP}_6^{[3]}$  provides very similar results as $\mathcal{S}_3$ (in fact, the curves in Figure 
 \ref{fig:effic} almost coincide)},
 this is not the case for the new schemes of orders 4 and 5,
 essentially due to the  much reduced number of exponentials involved. The improvement is even more remarkable in the case of the 6th-order 
 method $\mathcal{PCP}_{26}^{[6]}$, which
 in fact provides the most efficient approximation. This method should be compared with the 6th-order scheme labelled $\mathcal{G}_6$, 
 which is obtained by applying the recursion of \cite{chen22epf}
 to our scheme $\mathcal{NCP}_{10}^{[4]}$ (involving a total of 98 exponentials). In all cases, the leftmost point along the line corresponds to $n=1$.
 
 To discard the possible effect of the low dimensionality ($2 \times 2$) and the particular structure (skew-Her\-mi\-tian) of the matrices involved, we repeat the same experiment,
 but this time by constructing two $16 \times 16$ real matrices $\tilde{A}$, $\tilde{B}$ with elements randomly generated from a normal distribution and then taking 
 $A = \tilde{A} / \|\tilde{A}\|$, $B = \tilde{B} / \|\tilde{B}\|$. As before, we plot  in Figure \ref{fig:effic} (right) the efficiency diagram obtained by the methods of Table \ref{table.3} in comparison with the most efficient approximation designed in \cite{chen22epf}.
 We observe the same behavior as for Pauli matrices, with the new 6th-order method $\mathcal{PCP}_{26}^{[6]}$ showing the best performance.

\begin{figure}[!ht]
  \begin{center}
      \includegraphics[scale=0.48]{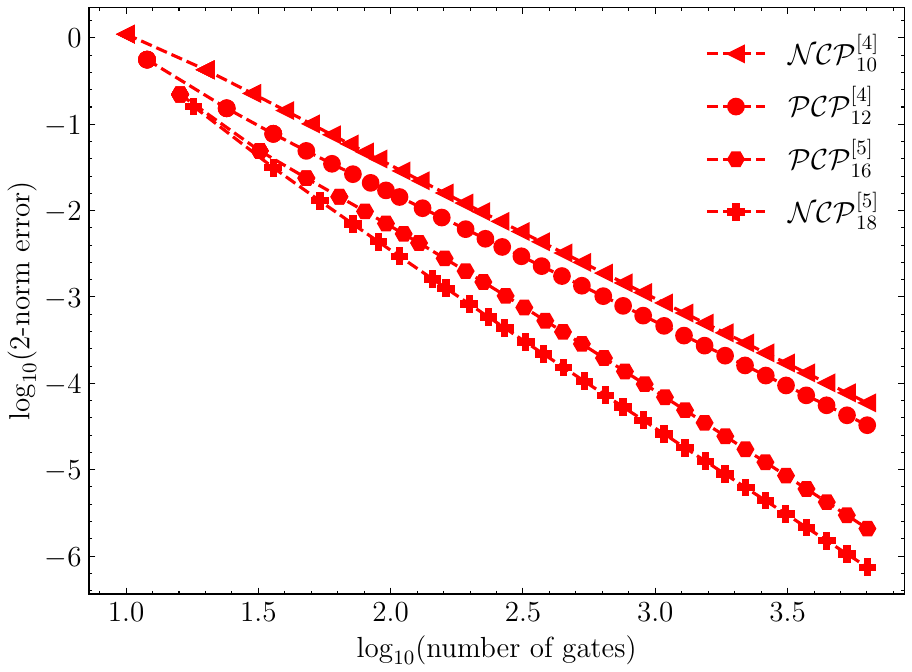}
      \includegraphics[scale=0.48]{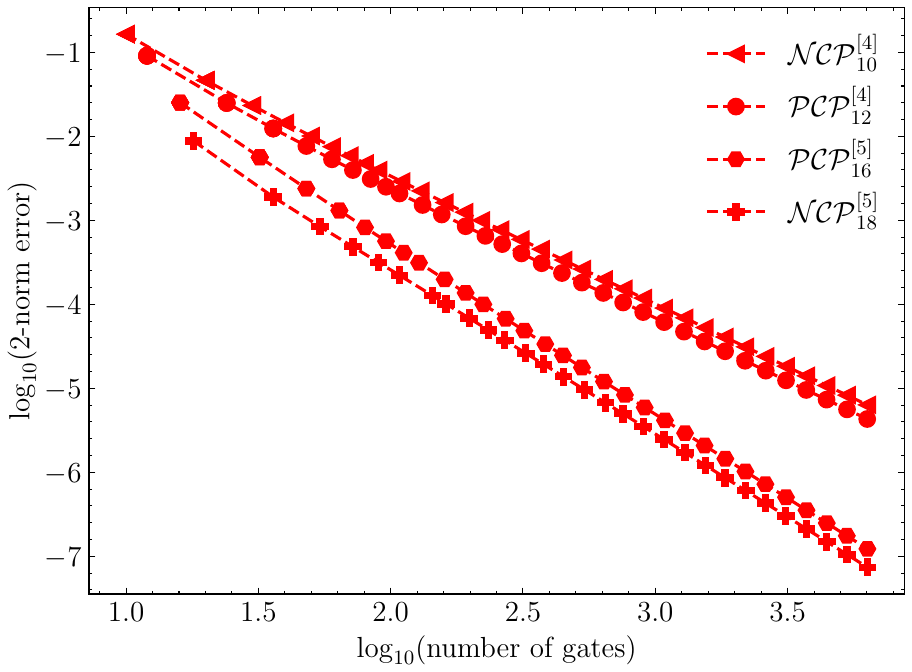}
    \caption{\small{Error in the approximation of $\e^{[A,B]}$ obtained by the new product formulas proposed in this work when
    $A = -i \sigma_x$, $B = -i \sigma_z$ (left) and when $A$ and $B$ are $16 \times 16$ real matrices with random elements (right). Here we check the improvement of the
    optimized schemes of Table  \ref{table.4} with respect to the product formulas of Table  \ref{table.3}. }}
    \label{fig:effic_improv}
  \end{center}
\end{figure}

Figure \ref{fig:effic_improv} illustrates, on the same examples, the behavior of the optimized methods of order 4 and 5 whose coefficients are collected in Table \ref{table.4}.
Notice that, even if the computational cost per step increases, their overall efficiency has improved in both cases.

\begin{figure}[!ht]
  \begin{center}
      \includegraphics[scale=0.48]{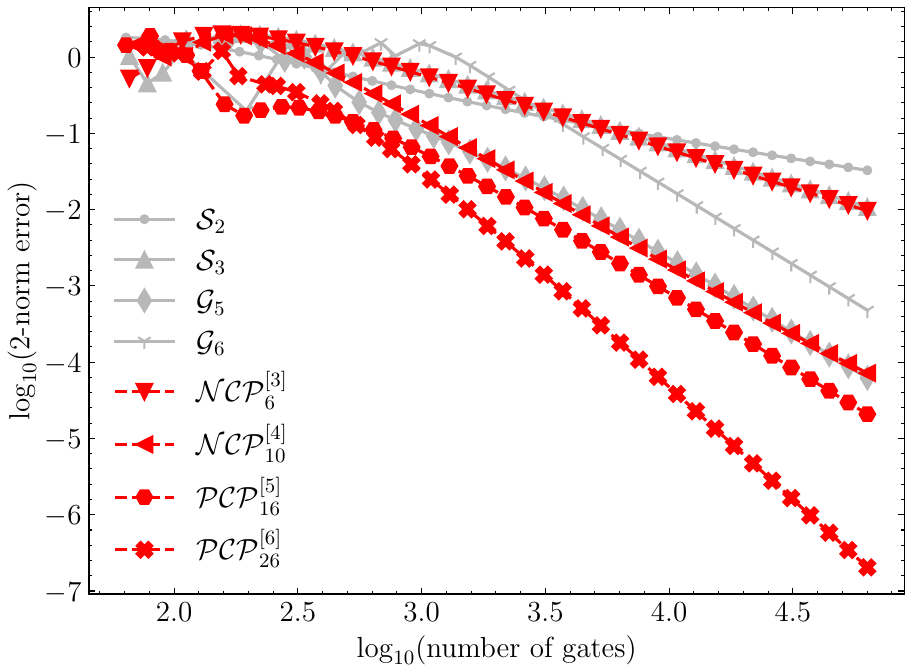}
      \includegraphics[scale=0.48]{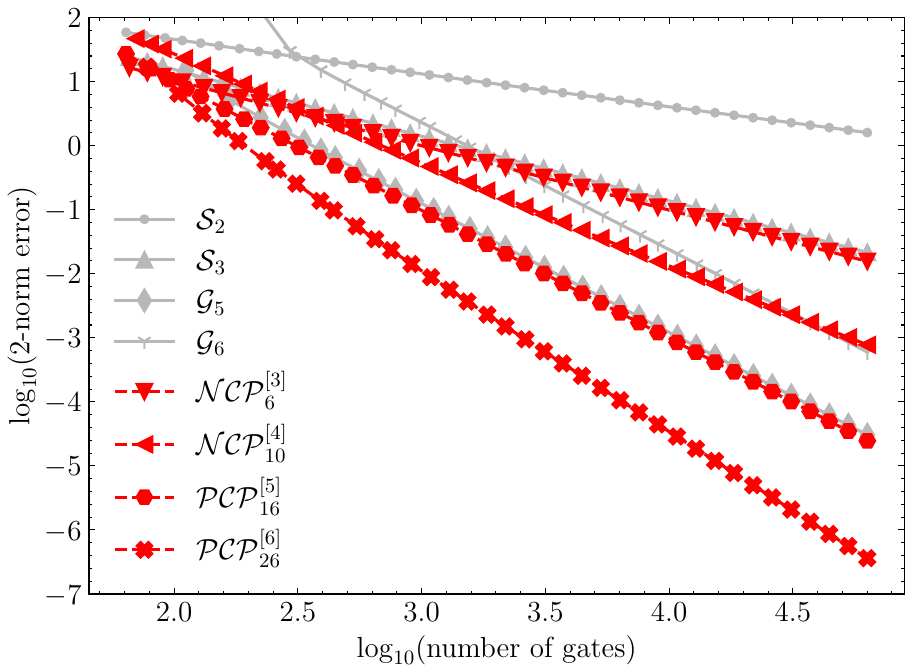}
    \caption{\small{Same as Figure \ref{fig:effic}, but now when approximating $\e^{t [A,B]}$ for a larger value $t=10$, with $A = -i \sigma_x$, $B = -i \sigma_z$. Notice the improved efficiency of the new
    6th-order approximation (in red), especially with respect to previous schemes (in gray).}}
    \label{fig:ef_10}
  \end{center}
\end{figure}

We can also consider values $t > 1$ to check how the different schemes perform in the long run. To this end, we take $t=10$ and compute accordingly
\[
    \big\| U(10/\sqrt{n})^n - \e^{10 [A,B]} \big\|_2.
\]
 The corresponding efficiency diagrams for Pauli (left) and random $16 \times 16$ matrices (right) are shown in Figure \ref{fig:ef_10}.

\begin{figure}
  \begin{center}
        \includegraphics[width=8cm]{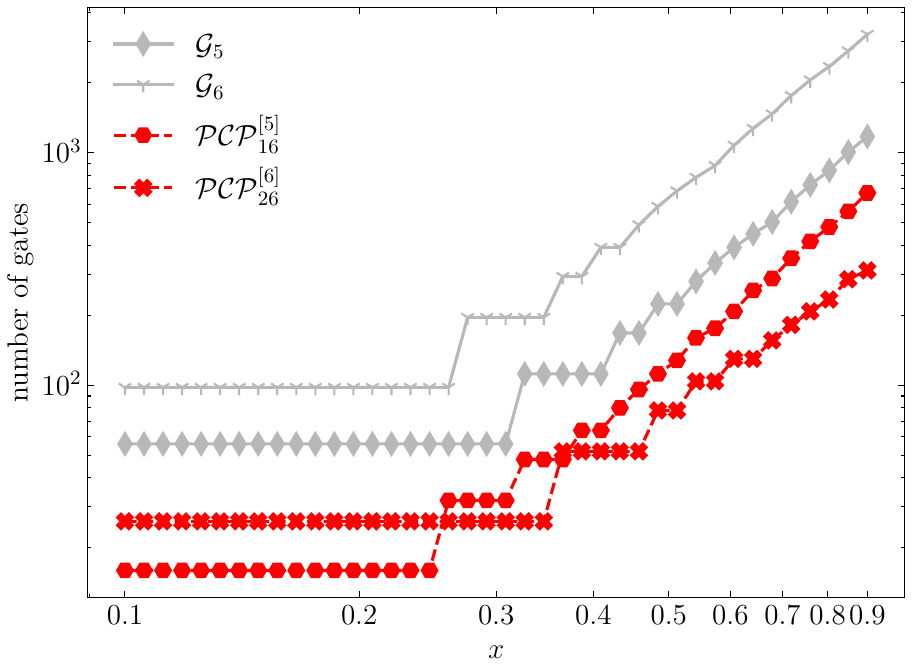}
       \includegraphics[width=8cm]{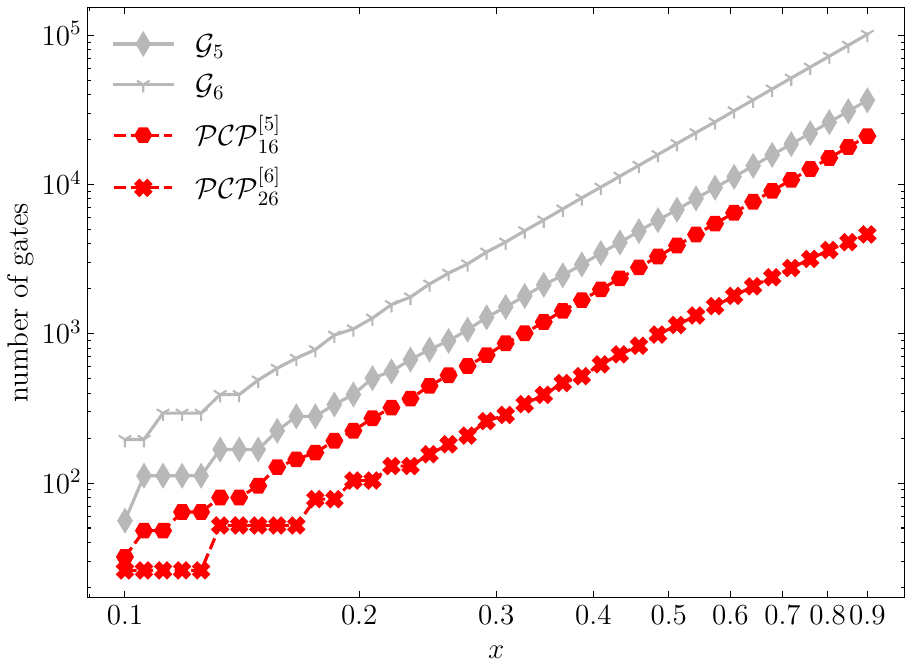}
    \caption{\small{Number of elementary exponentials (gates) required by each product formula to approximate $\e^{x^2 [-i \sigma_x, -i \sigma_z]}$ with error smaller than
    $10^{-4}$ (left) and $10^{-7}$ (right). This graph should be compared with Fig. 4 in \cite{chen22epf}: $\mathcal{G}_5$ (involving 56 exponentials) are depicted there, whereas 
    $\mathcal{G}_6$ (involving 98 exponentials) is obtained by applying the recursion of  \cite{chen22epf} to our 4th-order scheme $\mathcal{NCP}_{10}^{[4]}$. Again, the new approximations provide better results for all values of $x$ considered.}}
    \label{fig_accuracy}
  \end{center}
\end{figure}

In our next experiment, 
we determine the number of elementary exponentials  $n \cdot s$ required for each scheme $U(t)$ so that
the error for different values of $x$ in $\e^{x^2 [A,B]}$, namely
\begin{equation} \label{err_2}
   \big\| U(x/\sqrt{n})^n - \e^{x^2[A,B]} \big\|_2
\end{equation}
is smaller than a prescribed tolerance. 
 We choose again $A = -i \sigma_x$, $B = -i \sigma_z$, and $x \in [0.1, 0.9]$. The corresponding results
are depicted in Figure \ref{fig_accuracy} for a tolerance $10^{-4}$ (left) and $10^{-7}$ (right). The improvement with respect to the most efficient
scheme presented in \cite{chen22epf} is quite remarkable, even for small values of $x$.

\section{Extensions}
\label{sec.5}

Not only the exponential of a commutator can be approximated by products of exponentials of the elementary operators involved, but in fact the same technique
can be applied in principle to any Lie polynomial in $A$ and $B$. In this section we illustrate the procedure by considering particular cases.

\subsection{Approximating the exponential of $F = t (A + B) + t^2 R \, [A,B]$}

Suppose one is interested in computing an approximation to the exponential of both the sum and commutator of $A$, $B$, namely, $F = t (A + B) + t^2 R \, [A,B]$, for
an arbitrary $R\in\mathbb{R}$. A particular application arises in the simulation of counter-adiabatic time evolution of two-qubit Hamiltonians \cite{chen22epf}. 

Clearly, the coefficients of an approximation of the form \eqref{BA} of order $r$, 
\[
  \e^{c_0 t R B} \, \e^{c_1 t R A} \, \cdots \, \e^{c_{2n} t R B} \, \e^{c_{2n+1} t R A} = \e^F + \mathcal{O}(t^{r+1})
\]  
have to verify
\[
  w_{1,1} = w_{1,2} = 1, \qquad w_{2,1} = R^2, \qquad w_{j, \ell} = 0, \qquad 3 \le j \le r; \quad \ell = 1, \ldots, \dim(\cL_j).
\]  
Thus, we get an approximation of order 2 with only 3 exponentials:
\[
  \Phi_3^{[2]}(t) =\e^{c_0tRB} \, \e^{c_1tRA} \, \e^{c_2tRB}, \qquad\qquad
 c_0=- \frac{2 R^{2} - 1}{2 R}, \quad  c_1=\frac{1}{R}, \quad  c_2=\frac{2 R^{2} + 1}{2 R},
\]
whereas the inclusion of a free parameter $c_3$ allows us to reduce the effective error:
\[
   \Phi_4^{[2]}(t)=\e^{c_0tRB} \, \e^{c_1tRA} \, \e^{c_2tRB} \, \e^{c_3tRA},
\]
with 
\[
  c_0=\frac{2 R^{2} + 2 R c_{3} - 1}{2 R \left(R c_{3} - 1\right)}, \quad  c_1=- \frac{R c_{3} - 1}{R}, \quad  c_2=- \frac{2 R^{2} + 1}{2 R \left(R c_{3} - 1\right)}.
\]
Moreover, in this case we recover the scheme $U_{2,2}(t)$ of \eqref{or.2} in the limit $R \rightarrow \infty$ with $c_3 = -1$.

By following the same strategy, the following 3rd-order scheme can be obtained:
\[
  \Phi_5^{[3]}(t)=\e^{c_0tRB} \, \e^{c_1tRA} \, \e^{c_2tRB} \, \e^{c_3tRA} \, \e^{c_4tRB},
\]
with coefficients
\[
\begin{aligned}
  & c_0 =\frac{1}{R} \left(3R^4-R^2\mp \Delta +\frac{1}{4}\right),\quad  c_1=\frac{6R^4\pm 2\Delta-\frac{1}{2}}{R(12R^4-1)},\quad c_2=\frac{1}{2R}- 6R^3,\\
  & c_3 =\frac{6R^4\mp 2\Delta-\frac{1}{2}}{R(12R^4-1)},\qquad c_4=\frac{1}{R} \left(3R^4-R^2\pm \Delta +\frac{1}{4}\right), \\
  & \qquad  \Delta = \frac{\sqrt{3(12R^4-1)(36R^4+1)}}{12}.
 \end{aligned}
 \] 
It is worth noticing that, whereas 6 exponentials are necessary to approximate $\e^{t^2 [A,B]}$ up to order 3 (cf. Section \ref{sec.3}), with only 5 we are able to reproduce
$\e^F$ in general, in contrast with the treatment in \cite{chen22epf}.

\subsection{Approximating  $\e^{t^3 [A,[A,B]]}$}

In some problems arising in 
quantum computing it is desirable to suppress the effect of undesirable interactions in coupled spin systems by using composite pulse sequences to generate
an effective Hamiltonian for which the nested commutator $[A,[A,B]]$ is the dominant term \cite{borneman12pit}. The evolution due to the effective Hamiltonian can itself 
be approximated by a product of
exponentials of $A$ and $B$. This is also the case in classical control problems, where $[A,[A,B]]$ corresponds to a vector field along which the trajectory has to determined.

A straightforward calculation shows that
\begin{equation} \label{fap}
  \e^{t A} \, \e^{t B} \, \e^{-t A} \, \e^{-t B} \, \e^{-t A} \, \e^{t B} \, \e^{t A} \, \e^{-t B} = \e^{t^3 [A,[A,B]]} + \mathcal{O}(t^4),
\end{equation}
whereas, if the identity $[A,[B,[B,A]]] = 0$ holds, then the error of this approximation is of order $\mathcal{O}(t^5)$ \cite{childs13pff}. By applying the same strategy as in section
\ref{sec.3} it is possible to construct a left-right palindromic composition of order 4 involving only 9 exponentials,
\begin{equation} \label{aor4}
  U(t) = \e^{d_0 t B} \, \e^{d_1 t A} \, \e^{d_2 t B} \, \e^{d_3 t A} \, \e^{d_4 t B} \, \e^{d_3 t A} \, \e^{d_2 t B} \, \e^{d_1 t A} \, \e^{d_0 t B} = \e^{t^3 [A,[A,B]]} + \mathcal{O}(t^5).
\end{equation}
In fact, this composition has essentially the same computational cost as \eqref{fap} when it is used iteratively (since the last exponential can be concatenated with the first one at the next step). The analysis shows that
there is a free parameter in \eqref{aor4}, so that the sequence of coefficients reads
\[
  (d_0, d_1, d_2, d_3, d_4) =  \left(-\frac{d_2}{2}, \pm\frac{1}{\sqrt{d_2}}, d_2, \mp\frac{1}{\sqrt{d_2}}, -d_2 \right)
\]    
and the minimum effective error is achieved with $d_2 = \left((\sqrt{1346} - 36)/25 \right)^{1/3}$. By following the same approach, an approximation of order 6 can be
obtained with 21 exponentials.
On the other hand, the composition 
\[
  U(t) = \e^{d_0 t B} \, \e^{d_1 t A} \, \e^{d_2 t B} \, \e^{d_3 t A} \, \e^{d_4 t B}, 
\]
involving only 5 exponentials verifies
\[
   U(t) = \exp \big( t(A+B) + t^2 [A,B] + t^3 [A,[A,B]] \big) + \mathcal{O}(t^4)
\]   
when the coefficients are given by
\[
 d_0 = -\frac{3}{4} + \frac{\alpha}{4}, \quad d_1 = \frac{1}{2} + \frac{\alpha}{2}, \quad d_2 = \frac{1}{2}, \quad d_3 = \frac{1}{2} - \frac{\alpha}{2}, \quad
 d_4 = \frac{5}{4} - \frac{\alpha}{4}
\]   
and $\alpha = \sqrt{47/3}$.   
On the other hand, it is clear that the same product formulas can be used to approximate $\e^{t^3 [B,[B,A]]}$: we just have to interchange the role of $A$ and $B$ in the corresponding 
expressions. In fact, they can also be used recursively to construct approximations of nested commutators involving more operators, although they involve many more exponentials than
the minimum required. 

To illustrate this feature, a product formula of order 4 for $\e^{t^4 [A,[A,[A,B]]]}$ can be obtained as follows. We write
\[
   \e^{t^4 [A,[A,[A,B]]]} = \e^{t^2 [A, D]}, \qquad \mbox{ with } \qquad D = t^2 [A,[A,B]].
\]
Then, $\e^{t^2 [A, D]}$ is approximated by formula $\mathcal{NCP}_{10}^{[4]}$, 
\[
  \e^{c_0 t D} \, \e^{c_1 t A} \, \e^{c_2 t D} \, \e^{c_3 t A} \, \e^{c_4 t D} \, \e^{c_4 t A} \, \e^{c_3 t D} \, \e^{c_2 t A} \e^{c_1 t D} \, \e^{c_0 t A} 
\]
and finally we replace $\e^{c_j t D} = \e^{c_j t^3 [A,[A,B]]}$ in this formula by \eqref{aor4}. The final approximation of order 4 thus involves 50 elementary exponentials.
This number can be reduced by considering a palindromic sequence starting with $A$ to approximate $\e^{t^3 [A,[A,B]]}$.  

The previous formulae can be used in principle to carry out Hamiltonian simulation when the Hamiltonian includes the so-called modified potentials,
where some commutator terms are used to correct the error. This class of systems have been used in both the geometric integration and quantum literature 
\cite{bagherimehrab24faq}.

\subsection{Simulating the Zassenhaus formula}

Commutators also appear in a natural way in the Zassenhaus formula \cite{magnus54ote}. This can be described as the dual Baker--Campbell--Hausdorff formula, and expresses the exponential
of the sum of two non-commuting operators into an (in general infinite) product of exponentials operators involving linear combinations of nested commutators. Specifically,
\[
  \e^{t (A +B)} = \e^{t A} \, \e^{t B} \, \e^{t^2 C_2(A,B)} \, \e^{t^3 C_3(A,B)} \, \cdots \, \e^{t^n C_n(A,B)} \, \ldots,
\]
where $C_n(A,B)$ is a homogeneous Lie polynomial in $A$ and $B$ of degree $n$: a linear combination with rational coefficients of nested commutators involving
$n$ operators $A$ and $B$. An efficient algorithm to generate the terms $C_n$ for any $n$ has been proposed in \cite{casas12eco}. Thus, in particular,
\[
  C_2(A,B) = -\frac{1}{2} [A,B], \qquad C_3(A,B) = \frac{1}{3} [B,[A,B]] + \frac{1}{6} [A,[A,B]].
\]
While non-optimal as a simulation method, implementing the Zassenhaus formula is illustrative of how developing efficient algorithms for evolving commutators may lead to more efficient quantum algorithms \cite{burdine24tso}. As an illustration, we will use expression \eqref{aor4} to approximate the symmetric version of the Zassenhaus formula  proposed in \cite{arnal17ots}. 

The symmetric Zassenhaus formula reads
\[
   \e^{t (A +B)} = \e^{\frac{t}{2} A} \, \e^{\frac{t}{2} B} \, \e^{t^3 D_3} \, \e^{t^5 D_5} \, \cdots \, \e^{t^5 D_5} \, \e^{t^3 D_3} \, \e^{\frac{t}{2} B} \,  \e^{\frac{t}{2} A},
\]
where, again, $D_n$ is a homogeneous Lie polynomial of degree $n$. Notice that only odd powers of $t$ appear in this expression.
In consequence, 
\[
   \e^{t (A +B)} = \e^{\frac{t}{2} A} \, \e^{\frac{t}{2} B} \, \e^{2 t^3 D_3} \,  \e^{\frac{t}{2} B} \,  \e^{\frac{t}{2} A} + \mathcal{O}(t^5),
\]
where 
\[
  D_3 \equiv D_{3,1} + D_{3,2}, \qquad \mbox{ and } \qquad D_{3,1} = \frac{1}{24} [A,[A,B]], \qquad D_{3,2} = \frac{1}{12} [B,[A,B]]. 
\]
Clearly, then, it also holds that 
\begin{equation} \label{sym_zass_1}
   \e^{t (A +B)} = \e^{\frac{t}{2} A} \, \e^{\frac{t}{2} B} \, \e^{t^3 D_{3,1}} \,  \e^{t^3 D_{3,2}} \, \e^{\frac{t}{2} B} \,  \e^{\frac{t}{2} A} + \mathcal{O}(t^5),
\end{equation}
and this is the formula we simulate with the approximation \eqref{aor4} (and the corresponding one by interchanging $A$ and $B$).

\begin{figure}
  \begin{center}
        \includegraphics[width=8cm]{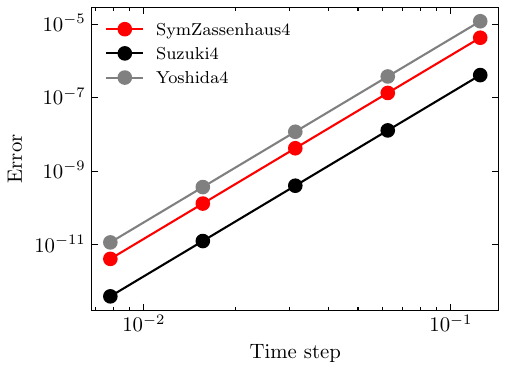}
       \includegraphics[width=8cm]{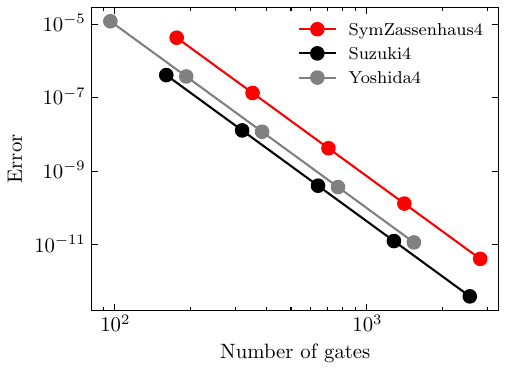}
    \caption{\small{Left: error vs time step for simulating $\e^{-i t (\sigma_x + \sigma_z)}$ using the triple-jump (Yoshida), Suzuki and symmetric Zassenhaus product formulas of order 4. All methods follow a 5th order power law for a single step. Right: Error vs cost for total simulation time $t = 1$.}}
    \label{fig_zassenhaus}
  \end{center}
\end{figure}

In Figure \ref{fig_zassenhaus} we compare the results achieved with the approximation \eqref{sym_zass_1} in comparison with the 4th-order Yoshida triple-jump
 \eqref{yoshida} and the Suzuki quintuple-jump composition \eqref{suzuki} with $k=2$. The left plot shows the error vs the step size, whereas the right diagram
 corresponds to the error vs the computational cost when $A = -i \sigma_x$, $B = -i \sigma_z$ and $t=1$. These graphs exhibit the correct scaling of the techniques presented here, although the Suzuki method shows a more favorable scaling, due to the smaller number of elementary exponentials required (20 
 instead of 22 for the symmetric Zassenhaus formula). In any case, we believe the methods presented here may find applicability in problems with a special structure where the implementation of specific commutators may allow product formulas to achieve a higher error reduction at a moderate cost. The code to generate
 these figures is available at \url{https://github.com/PabloAMC/exponentials_commutators/}.

\section{Concluding remarks}
\label{sec.6}

In this work we have obtained efficient approximations for the exponential of the commutator $\e^{t^2 [A,B]}$ of two arbitrary operators $A$ and $B$ as compositions
 of exponentials of the operators involved. 
 Previous analyses of this same problem show that it is in fact possible to get arbitrarily high-order approximations by considering
 recursive formulas starting from a basic low-order scheme, in a similar vein to  the well known triple \eqref{yoshida} and quintuple jump \eqref{suzuki} formulas do for 
 approximating $\e^{t (A + B)}$.
 Although these recursive formulas allow one to get explicit upper bounds for the error committed and 
 exponentials required, this number is exceedingly large, even for moderate orders of approximation, and so their application to solve problems in practice is questionable at least. This
 is the case in particular in quantum computing, where the action of $\e^{t A}$ and $\e^{t B}$ is simulated by a quantum circuit. In that setting it is clearly relevant to reduce as much as
 possible the number of exponentials required to achieve a certain accuracy in the approximation. 
 
 Here, by applying the same techniques as in the construction of splitting methods for the time integration of differential equations, and considering particular patterns in the
 distribution of the exponentials, we have obtained optimized product formulas
 of orders 4, 5 and 6 with a much reduced computational cost (only two more exponentials than the theoretical minimum). The numerical examples we gather clearly show
 how this reduction greatly contributes to the improvement in the efficiency. The new methods could then be applied in quantum simulations where it is essential to have a
number of gates as small as possible whereas still having good accuracy. Other possible applications include problems in classical and quantum control, 
the computation of the successive terms in the Zassenhaus formula 
\cite{casas12eco} and
its continuous symmetric analogue \cite{bader14eaf}.
 
 Although we have limited our analysis up to order 6, it
 is clear that similar calculations can be carried out for higher orders if necessary. In any case, the new methods can be used as basic schemes for previously generated 
recursive procedures. One should take into account, however, that although recursive product formulas are very valuable, for practical computations where a given order of accuracy is
sought, direct formulas as those presented here are clearly preferred, given their reduced computational cost.

\subsection*{Acknowledgements} 
 This work has been
funded by Ministerio de Ciencia e Innovaci\'on (Spain) through project PID2022-136585NB-C21, 
MCIN/AEI/10.13039/501100011033/FEDER, UE, and also by Generalitat Valenciana (Spain) through project CIAICO/2021/180.
The authors wish to thank S. Blanes for several comments and remarks.

\subsection*{Declarations}
All authors have contributed equally to this work. The authors declare no conflict of interest.

\bibliographystyle{siam}

\end{document}